\documentclass[conference]{IEEEtran}
\usepackage{cite}
\usepackage{amsmath,amssymb,amsfonts}
\usepackage{algorithmic}
\usepackage{graphicx}
\usepackage{textcomp}
\usepackage{xcolor}
\def\BibTeX{{\rm B\kern-.05em{\sc i\kern-.025em b}\kern-.08em
    T\kern-.1667em\lower.7ex\hbox{E}\kern-.125emX}}

\usepackage{xspace}
\usepackage{multirow}
\usepackage[draft]{hyperref}
\usepackage{makecell, tabularx}
\newcolumntype{C}{>{\centering\arraybackslash}X} %

\definecolor{bl}{RGB}{0, 87, 183} %
\definecolor{gelb}{RGB}{175, 175, 175} %
\definecolor{blau}{RGB}{0, 0, 0} %

\newcommand{\CS}{$\textit{CS}$\xspace}
\newcommand{\dCS}{$\Delta\textit{CS}$\xspace}
\newcommand{\Rtwo}{$R_2^2$\xspace}
\newcommand{\dRtwo}{$\Delta R_2^{2}$\xspace}
\newcommand{\Rp}{$R_p^2$\xspace}
\newcommand{\CE}{$H$\xspace}
\newcommand{\dCE}{$\Delta H$\xspace}
\newcommand{\dERz}{$\Delta\textit{ER}^\mathtt{0}$\xspace}
\newcommand{\dERnz}{$\Delta\textit{ER}^\mathtt{n0}$\xspace}
\newcommand{\dFMed}{$\Delta\textit{FM}_\textit{ed}$\xspace}
\newcommand{\dFMr}{$\Delta\textit{FM}_r$\xspace}
\newcommand{\dFM}{$\Delta\textit{FM}$\xspace}
\newcommand{\K}{$K$\xspace}
\newcommand{\KL}{$\textit{KL}$\xspace}
\newcommand{\dKL}{$\Delta\textit{KL}$\xspace}
\newcommand{\LV}{$\textit{LV}$\xspace}
\newcommand{\MIs}{$\textit{MI}_s$\xspace}
\newcommand{\MIi}{$\textit{MI}^\mathtt{i}$\xspace}
\newcommand{\MIik}{$\textit{MI}^\mathtt{i}(\textrm{NN})$\xspace}
\newcommand{\MIm}{$\textit{MI}^\mathtt{m}$\xspace}
\newcommand{\dMIb}{$\Delta\textit{MI}^\mathtt{b}$\xspace}
\newcommand{\dMIs}{$\Delta\textit{MI}_s$\xspace}
\newcommand{\dMIi}{$\Delta\textit{MI}^\mathtt{i}$\xspace}
\newcommand{\RUr}{$\textit{RU}^\mathtt{r}$\xspace}
\newcommand{\RUw}{$\textit{RU}^\mathtt{w}$\xspace}
\newcommand{\TVD}{$\textit{TVD}$\xspace}

\newcommand{\PM}{\textit{PM}\xspace}

\newcommand{\scor}{\textit{cor}\xspace}
\newcommand{\pval}{\textit{pv}\xspace}

\newcommand{\prm}{\textit{pm}\xspace}

\begin{document}
\bstctlcite{IEEEexample:BSTcontrol}

\title{Evaluating Privacy Measures for Load Hiding}

\author{\IEEEauthorblockN{Vadim Arzamasov}
\IEEEauthorblockA{
\textit{Karlsruhe Institute of Technology}\\
Karlsruhe, Germany \\
vadim.arzamasov@kit.edu
}
\and
\IEEEauthorblockN{Klemens B\"ohm}
\IEEEauthorblockA{
\textit{Karlsruhe Institute of Technology}\\
Karlsruhe, Germany \\
klemens.boehm@kit.edu
}
}

\maketitle

\begin{abstract}
In smart grids, the use of smart meters to measure electricity consumption at a household level raises privacy concerns. 
To address them, researchers have designed various load hiding algorithms that manipulate the electricity consumption measured. 
To compare how well these algorithms preserve privacy, various privacy measures have been proposed. 
However, there currently is no consensus on which privacy measure is most appropriate to use. 
In this study, we aim to identify the most effective privacy measure(s) for load hiding algorithms. 
We have crafted a series of experiments to assess the effectiveness of these measures. 
found 20 of the 25 measures studied to be ineffective. 
Next, focused on the well-known ``appliance usage'' secret, we have designed synthetic data to find the measure that best deals with this secret. 
We observe that such a measure, a variant of mutual information, actually exists. 
\end{abstract}


\section{Introduction}

The shift towards renewable energy has prompted the use of innovative electrical grid designs, such as smart grids. 
Smart grids leverage smart meters to enhance efficiency, reliability, and sustainability by continuously measuring and transmitting electricity consumption data (often called \emph{load profiles}) with high accuracy. 
However, privacy concerns arise when measuring consumption at the household level. 
To address this, various load hiding algorithms have been proposed.
To assess the effectiveness of these algorithms in preserving privacy, literature features various \emph{privacy measures}. 
Table~\ref{tab:general-measures-references} lists some of them. 
The choice of measure can significantly impact the comparison of load-hiding algorithms~\cite{DBLP:conf/eenergy/ArzamasovSKBN20}. 
But there is no consensus on which measure is most appropriate.

\begin{table}[htbp]
	\caption{Popular smart meter privacy measures~\cite{DBLP:conf/eenergy/ArzamasovSKBN20}}
	\begin{center}
		\begin{tabular}{|c|c|c|}
			\hline
			\textbf{Category} & \textbf{Measure} & \textbf{Reference} \\
			\hline
			\multirow{2}{*}{Cluster similarity} & \CS & \cite{Kalogridis2011,Kalogridis2011c} \\
			& \dCS & \cite{Kalogridis2011c,Kalogridis2010a} \\ 
			\hline
			\multirow{3}{*}{Coefficient of determination} & \Rtwo & \cite{Kalogridis2011c} \\ 
			& \dRtwo & \cite{Kalogridis2011,Kalogridis2011c,Kalogridis2010a} \\
			& \Rp & \cite{Chen2013} \\
			\hline
			\multirow{2}{*}{Conditional entropy} & \CE & \cite{Yao2015,Castro2009} \\
			& \dCE & \cite{Kalogridis2010a} \\ 
			\hline
			\multirow{2}{*}{Entropy ratio} & \dERz & \cite{McLaughlin2011} \\ 
			& \dERnz & \cite{McLaughlin2011} \\
			\hline
			\multirow{3}{*}{Feature mass} & \dFMed & \cite{Zhao2014a} \\
			& \dFMr & \cite{McLaughlin2011} \\
			& \dFM & \cite{Yang2014} \\
			\hline			
			K-divergence & \K & \cite{Kalogridis2011b} \\
			\hline
			\multirow{2}{*}{KL divergence} & \KL & \cite{Zhu2017} \\ 
			& \dKL & \cite{Kalogridis2011a,Kalogridis2010a,Yang2014OptimalMeters} \\
			\hline
			Load variance & \LV & \cite{Tan2017a} \\ 
			\hline
			\multirow{5}{*}{Mutual information$^{\mathrm{a}}$} & \MIs & \cite{Tan2017a} \\
			& \MIi & \cite{Chin2018, Giaconi2015, Zhu2017, Gomez-vilardebo2013, Zhao2014a, Chin2016, Erdogdu2015Privacy-UtilityData, Giaconi2017a}\\
			& \MIm & \cite{Chin2018} \\
			& \dMIb & \cite{Yang2012b} \\
			& \dMIs & \cite{Yang2012b} \\
			& \dMIi & \cite{Zhao2014a,Yang2012b} \\
			\hline
			\multirow{2}{*}{Removed uncertainty} & \RUr & \cite{Laforet2015} \\ 
			& \RUw & \cite{Laforet2015} \\
			\hline
			Total variation distance & \TVD &\cite{Zhu2017,Kalogridis2011}, \cite{Backes2014}$^{\mathrm{b}}$ \\ \hline
			\multicolumn{3}{l}{$^{\mathrm{a}}$Examples of other work that use or mention mutual information as} \\ 
			\multicolumn{3}{l}{a privacy measure for load hiding methods are~\cite{Arrieta2018SmartChannel,Koo,Li2015,Rajagopalan2011a,Sankar2013,Castro2009,Tan2013a,Varodayan2011a}} \\
			\multicolumn{3}{l}{$^{\mathrm{b}}$It is not clear from the description whether time differences are taken.}
		\end{tabular}
		\label{tab:general-measures-references}
	\end{center}
\end{table}

The privacy measures proposed for load hiding algorithms seem reasonable and are supported by theory. 
Determining the most effective measures thus requires experimentation, which can be challenging. 
This is because generating load profiles with known privacy levels is often infeasible. 
This calls for experiments that reveal how the various measures behave in situations with a known outcome and discard the ones whose results are not in line with this.
For example, one can create a setting where the privacy level gradually increases and require the measure to indicate this.

Our contributions are as follows.

\subsubsection{Compilation of Secrets from the Literature}
A privacy measure should assess the ability of load-hiding algorithms to conceal private information, or \emph{secrets}, in a load profile. 
In Section~\ref{section:data}, we have compiled the types of secrets featured in the literature. 
We observe that, so far, the connection between the ability of load-hiding algorithms to conceal secrets and privacy measures has not been studied systematically.

\subsubsection{Generic Requirements}
Our study aims to find the most effective privacy measure(s) for load hiding methods. 
In Section~\ref{section:methodology}, we propose natural requirements that privacy measures should meet and test the 25 measures listed in Table~\ref{tab:general-measures-references} against them. 
Only five measures pass at least six of seven tests.

\subsubsection{Consistency of Measures} 
Section~\ref{section:methodology_consistency} compares the behavior of different privacy measures. 
We observe moderate consistency: out of the five measures that satisfy most requirements, only two behave similarly and thus are interchangeable.

\subsubsection{Hiding a Secret} 
In Section~\ref{section:case_study}, we evaluate the ability of privacy measures to quantify the effectiveness of a load hiding algorithm in concealing a specific secret. 
We use synthetic data to this end.
Our analysis reveals that some measures are inadequate in this regard, and we discuss their limitations.

\subsubsection{Estimator Choice}
For two measures, \MIi and \CE, which have performed well so far, various \emph{estimators} to calculate them exist. 
Previous studies solely relied on the conventional histogram estimator and have not assessed its impact. 
In Section~\ref{section:estimator}, we compare this estimator with a nearest-neighbor-based alternative.
We observe that the histogram estimator can be inaccurate, and the alternative tends to perform better.

\subsubsection{Reproducibility}
We share the code of our experiments\footnote{\url{https://anonymous.4open.science/r/privacy_measures-824B}}. The repository also includes detailed descriptions of the experimental setting and auxiliary experiment results, which we mention in the article but don't include for brevity.

\section{Notation}
\label{section:notations}

The original load profile, the \emph{user load}, is $x=(x_{1}, \dots, x_T)$.
It consists of energy values in kWh consumed in the period $[t-1,t)$ or power in kW needed at time $t$ measured at equidistant times $t\in\{1, \ldots, T\}$. 
A modified load profile, \emph{grid load}, $y=(y_{1}, \dots, y_{T})$ is the result of applying a load hiding method to $x$.
When experimenting with multiple user loads, we use superscripts to differentiate between them, such as $x^i$ for $i=1,\dots,N$. 
Some privacy measures
use first differences $\Delta x=(\Delta x_{2}, \dots, \Delta x_T)$, where $\Delta x_{t}=x_{t}-x_{t-1}$, $t\in{2,\ldots,T}$ instead of $x$.
Similarly they replace $y$ with $\Delta y$. 
Let $u(k) = (u_1(k), \dots, u_T(k))$ be the noise profile; 
its elements $u_t(k)$ are independently and uniformly distributed in the interval $[-k,k]$. 
Our approach relies on the $\oplus$ operation: $x\oplus u(k)=y(k)$, so that $y_t(k)\ge0$ and $\sum_t^T x_t=\sum_t^T y_t(k)$.

A privacy measure $\PM(x,y,\theta)$ is a function that quantifies privacy resulting from substituting $x$ with $y$. 
$\theta$ is its hyperparameter(s), 
for example, the number of bins $n$ to discretize load profiles in the histogram estimator of mutual information. 
We often use the default values of $\theta$ from the literature and omit~$\theta$ to simplify notation. 
A greater value of $\PM(x,y)$ indicates higher privacy, unless stated otherwise.

The second column of Table~\ref{tab:general-measures-references} lists the privacy measures used in our experiments. 
We largely adhere to the notation from~\cite{DBLP:conf/eenergy/ArzamasovSKBN20}, but add the symbol $\Delta$ to indicate that the measure takes the differences of load profiles, denoted as $\Delta x$ and $\Delta y$, as inputs. 
The symbols $\mathtt{0}$ and $\mathtt{n0}$ in \dERz and \dERnz indicate whether calculation of the entropy ratio includes zero values or not. 
\RUr and \RUw refer to the measures of removed uncertainty using regression or wavelet filtering. 
For more information, please refer to the citations listed in Table~\ref{tab:general-measures-references}.

\section{Data}
\label{section:data}

We compare measures using real data. We create a shortlist of measures to investigate by removing those that do not meet the requirements identified in this study. We use synthetic data to evaluate the effectiveness of measures in detecting information on appliance usage hidden in load profiles.

\subsection{Real Data}

We use two popular data sources for our experiments: the Smart* dataset\footnote{\url{https://traces.cs.umass.edu/index.php/Smart/Smart}; the `Apartment dataset'} contains power measurements for 114 single-family apartments; the CER dataset~\cite{CER_Smart_Metering_Project} contains energy consumption data from over 5000 Irish homes and businesses. 
Smart* contains one measurement per minute, while CER measures every 30 minutes. Our experiments use 50 load profiles from each dataset, with Smart* profiles limited to one day, and CER profiles are four weeks long. This covers typical periods and frequencies used in the literature. To keep the paper concise, we present only the results for CER data when they are similar to Smart* data results.

\subsection{Synthetic Data}
\begin{table}[t]%
	\caption{Secrets literature deems detectable in load profiles.}
	\begin{center}
	\begin{tabularx}{\linewidth}{|c|@{}C@{}|}
		\hline
		\textbf{Secret} & \textbf{References} \\
		\hline
		Activities & \cite {Lisovich2008,Rajagopalan2011a,Koo,Laforet2016a,Yang2014,Buchmann,Yang2014OptimalMeters,Cavoukian2010SmartPrivacyConservation,McLaughlin2011}\\ \hline
		Age of appliances & \cite{Zhao2014a}\\ \hline
		\raisebox{-.5\height}{Appliances usage} & \cite{Yang2012b,Kalogridis2010a,Tan2017a,Avula2018,Lisovich2008,Tan2013a,Zhao2014a,Chin2016,Kalogridis2011,Kalogridis2011a,Backes2014,Efthymiou2010,Arrieta2018SmartChannel,Li2018SmartModels,Tudor2013AnalysisGrid,Castro2009,Cavoukian2010SmartPrivacyConservation,Giaconi2017a,Kessler2016,Varodayan2011a,Chin2018,Farokhi2017,FinsterPrivacy-awareSurvey,Chen2013,Buchmann,Rouf2004NeighborhoodSystems,Giaconi2015,Zhu2017,Gomez-vilardebo2013}\\ \hline
		Beliefs & \cite{Lisovich2008,FinsterPrivacy-awareSurvey}\\ \hline
		Brands of appliances & \cite{Engel2013a,Zhao2014a} \\ \hline
		Employment status & \cite{Laforet2015}\\  \hline
		Habits & \cite{McLaughlin2011,Giaconi2017a,Laforet2015,Rajagopalan2011a,Eibl2015,Chin2018,FinsterPrivacy-awareSurvey,Zhao2014a,Chin2016}\\ \hline
		Illness or disability & \cite {Giaconi2017a,Chin2018}\\ \hline
		Location of occupants & \cite{Zhao2014a}\\ \hline
		Living condition & \cite {McLaughlin2011}\\ \hline
		Meals & \cite{McLaughlin2011,Tan2017a,Molina-Markham2010PrivateMeter,Eibl2015,Rouf2004NeighborhoodSystems,Cavoukian2010SmartPrivacyConservation}\\ \hline
		Number of occupants & \cite{Yang2012b,Zhao2014a,Lisovich2008,Laforet2015,Molina-Markham2010PrivateMeter,Rouf2004NeighborhoodSystems}\\ \hline
		Preferences & \cite{Lisovich2008,Chin2016}\\ \hline
		Relationship status & \cite{Laforet2016a}\\ \hline
		Social class & \cite{Laforet2016a}\\ \hline
		Sleep-wake cycles
		&\cite{Engel2013a,Yang2012b,Tan2017a,Koo,Molina-Markham2010PrivateMeter,Eibl2015,Rouf2004NeighborhoodSystems,Lisovich2008} \\ \hline
		TV channels & \cite{Giaconi2017a,Tan2013a,Laforet2015}\\ \hline
		\raisebox{-.5\height}{Vacancy patterns}
		&\cite{Engel2013a,Yang2012b,McLaughlin2011,Zhao2014a,Chin2016,Giaconi2017a,Lisovich2008,Kessler2016,Tan2013a,Rajagopalan2011a,Yang2014,Eibl2015,Arrieta2018SmartChannel,Li2018SmartModels,Chin2018,Farokhi2017,FinsterPrivacy-awareSurvey,Tudor2013AnalysisGrid,Rouf2004NeighborhoodSystems,Giaconi2015,Gomez-vilardebo2013,Yang2014OptimalMeters,Cavoukian2010SmartPrivacyConservation,Tan2017a} \\ \hline
		Vacations & \cite{FinsterPrivacy-awareSurvey}\\ \hline
		Working hours & \cite {Giaconi2017a,Koo,Chin2018,FinsterPrivacy-awareSurvey}\\
		\hline
	\end{tabularx}
	\label{tab:private_information}
\end{center}
\end{table}

\begin{figure}[t]
	\centering
	\includegraphics[width=0.363\columnwidth]{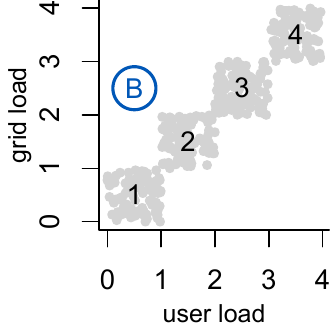}
	\quad
	\includegraphics[width=0.264\columnwidth]{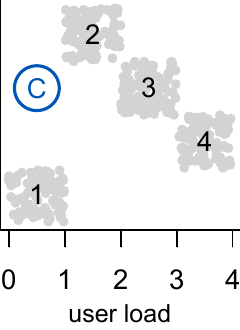}
	\quad
	\includegraphics[width=0.264\columnwidth]{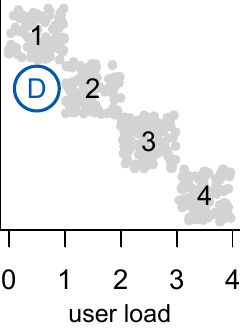}
	\caption{Examples of synthetic datasets}
	\label{fig:load_hiding_algorithms}
\end{figure}

Privacy measures aim to detect sensitive information, or secrets, removed from a load profile. Table~\ref{tab:private_information} shows examples of secrets. 
No systematic study has examined the relationship between effectiveness of load-hiding algorithms and privacy measures. 
Since covering all secrets is beyond the scope of one study, we focus on the appliances usage secret that is frequent in related work (see Table~\ref{tab:private_information}). %
We propose four algorithms that generate pairs of user load $x$ and grid load $y$ to study this secret. 
In some cases, $y$ conceals the secret, but not in others. 
Our study provides insights into using several measures and may serve as a blueprint for studying other secrets.

Consider a household with four electric devices numbered~1--4. Each device is active for six consecutive hours per day, with only one device on at a time. The power level of Device~1 is i.i.d. uniformly from [0,1], Devices~2, 3, and 4 draw power from [1,2], [2,3], and [3,4], respectively. 
Think of privacy-preserving algorithms, labeled A, B, C, and D, that manipulate the grid load. Algorithm A selects values i.i.d. uniformly from [0,4]. Algorithm B chooses a value from the interval corresponding to the currently active device, and Algorithms C and D swap some power ranges. Figure~\ref{fig:load_hiding_algorithms} shows scatter plots of user load versus grid load for Algorithms~B--D.

Algorithm A hides the ``appliances usage'' secret, while B does not, as it reveals the active device by revealing the actual power range. 
The effectiveness of Algorithms~C and D varies depending on the information available to the third party. If the algorithm in use is known, both C and D offer the same privacy as Algorithm~B. 
Otherwise, C may not be worse than Algorithm B since the ranges of user load and grid load coincide only for Devices 1 and 3. 
Similarly, Algorithm D is likely to be at least as good as B and C in this scenario. --
We will experiment with synthetic data in Section~\ref{section:case_study}.

\section{Natural Requirements}
\label{section:methodology}

\begin{table}[t]
	\caption{Privacy measure compliance with requirements on CER data.}
	\begin{center}
		\begin{tabular}{|c|c|c|c|c|c|c|c|}
			\hline
			\textbf{measure} & \textbf{AN} & \textbf{IN} & \textbf{C} & \textbf{SY} & \textbf{BP\textsubscript{1}} & \textbf{BP\textsubscript{2}} & \textbf{LP}\\ \hline
			\Rtwo & {\color{blau}$0.99$} & {\color{blau}$1$} & {\color{blau}$1$} & {\color{blau}$1$} & {\color{blau}$0.96$} & {\color{blau}$1$} & {\color{blau}$1$}\\ \hline
			\MIm & {\color{blau}$0.98$} & {\color{blau}$0.99$} & {\color{blau}$1$} & {\color{blau}$1$} & {\color{blau}$1$} & {\color{blau}$0.92$} & {\color{blau}$1$}\\ \hline
			\MIi & {\color{blau}$0.98$} & {\color{blau}$1$} & {\color{blau}$1$} & {\color{blau}$1$} & {\color{blau}$0.94$} & {\color{blau}$0.97$} & {\color{blau}$1$}\\ \hline
			\dRtwo & {\color{blau}$0.99$} & {\color{blau}$0.93$} & {\color{blau}$1$} & {\color{blau}$1$} & {\color{gelb}$0.86$} & {\color{blau}$0.94$} & {\color{blau}$1$}\\ \hline
			\CE & {\color{blau}$0.98$} & {\color{blau}$1$} & {\color{blau}$1$} & {\color{gelb}$12.1$} & {\color{blau}$1$} & {\color{blau}$0.97$} & {\color{blau}$1$}\\ \hline
			\dMIi & {\color{blau}$0.98$} & {\color{blau}$1$} & {\color{blau}$1$} & {\color{blau}$1$} & {\color{gelb}$0.84$} & {\color{gelb}$0.79$} & {\color{blau}$1$}\\ \hline
			\dMIb & {\color{blau}$0.92$} & {\color{gelb}$0.77$} & {\color{blau}$0.98$} & {\color{blau}$1$} & {\color{blau}$0.94$} & {\color{gelb}$0.88$} & {\color{blau}$1$}\\ \hline
			\TVD & {\color{gelb}$0.73$} & {\color{blau}$0.99$} & {\color{blau}$0.96$} & {\color{blau}$1$} & {\color{gelb}$0.32$} & {\color{gelb}$0.7$} & {\color{blau}$1$}\\ \hline
			\Rp & {\color{blau}$0.99$} & {\color{blau}$1$} & {\color{blau}$1$} & {\color{gelb}$0.81$} & {\color{gelb}$0$} & {\color{gelb}$0.84$} & {\color{blau}$1$}\\ \hline
			\dCE & {\color{blau}$0.98$} & {\color{blau}$1$} & {\color{blau}$1$} & {\color{gelb}$17.9$} & {\color{gelb}$0.8$} & {\color{gelb}$0.79$} & {\color{blau}$1$}\\ \hline
			\MIs & {\color{gelb}$0.13$} & {\color{blau}$1$} & {\color{blau}$0.98$} & {\color{blau}$1$} & {\color{blau}$1$} & {\color{gelb}$0$} & {\color{gelb}$0.88$}\\ \hline
			\dMIs & {\color{gelb}$-0.12$} & {\color{blau}$1$} & {\color{blau}$0.99$} & {\color{blau}$1$} & {\color{gelb}$0.6$} & {\color{gelb}$0$} & {\color{blau}$0.91$}\\ \hline
			\CS & {\color{gelb}$0.83$} & {\color{gelb}$0.81$} & {\color{gelb}$0.85$} & {\color{gelb}$2.3$} & {\color{blau}$1$} & {\color{blau}$0.95$} & {\color{blau}$1$}\\ \hline
			\K & {\color{gelb}$0.84$} & {\color{blau}$0.99$} & {\color{blau}$0.97$} & {\color{gelb}$1.3$} & {\color{gelb}$0.38$} & {\color{gelb}$0.75$} & {\color{blau}$1$}\\ \hline
			\dFMr & {\color{blau}$0.97$} & {\color{blau}$1$} & {\color{gelb}$0$} & {\color{gelb}$1.1$} & {\color{blau}$1$} & {\color{gelb}$0.44$} & {\color{gelb}$0.52$}\\ \hline
			\KL & {\color{gelb}$0.49$} & {\color{gelb}$0$} & {\color{gelb}$0.24$} & {\color{gelb}$0.76$} & {\color{blau}$1$} & {\color{blau}$1$} & {\color{blau}$1$}\\ \hline
			\dKL & {\color{gelb}$0.48$} & {\color{gelb}$0$} & {\color{gelb}$0.44$} & {\color{gelb}$2.9$} & {\color{blau}$1$} & {\color{blau}$1$} & {\color{blau}$1$}\\ \hline
			\dFM & {\color{gelb}$0.42$} & {\color{blau}$1$} & {\color{gelb}$0.86$} & {\color{gelb}$0$} & {\color{blau}$0.94$} & {\color{gelb}$0$} & {\color{gelb}$0.49$}\\ \hline
			\dERnz & {\color{gelb}$-0.67$} & {\color{blau}$1$} & {\color{gelb}$0.87$} & {\color{gelb}$1.3$} & {\color{blau}$1$} & {\color{gelb}$0$} & {\color{gelb}$0.48$}\\ \hline
			\LV & {\color{gelb}$-0.6$} & {\color{blau}$1$} & {\color{blau}$1$} & {\color{gelb}$0.41$} & {\color{gelb}$0$} & {\color{gelb}$0.57$} & {\color{gelb}$0.57$}\\ \hline
			\dCS & {\color{gelb}$0.63$} & {\color{gelb}$0.16$} & {\color{gelb}$0.84$} & {\color{gelb}$2.6$} & {\color{gelb}$0.76$} & {\color{gelb}$0.89$} & {\color{blau}$1$}\\ \hline
			\dFMed & {\color{gelb}$0.37$} & {\color{gelb}$0.1$} & {\color{gelb}$0.39$} & {\color{gelb}$15.4$} & {\color{gelb}$0.9$} & {\color{gelb}$0.89$} & {\color{blau}$1$}\\ \hline
			\dERz & {\color{gelb}$-0.88$} & {\color{blau}$1$} & {\color{gelb}$0.87$} & {\color{gelb}$1.3$} & {\color{gelb}$0.22$} & {\color{gelb}$0$} & {\color{gelb}$0.48$}\\ \hline
			\RUr & {\color{gelb}$-0.95$} & {\color{gelb}$0$} & {\color{gelb}$0.7$} & {\color{gelb}$1.3$} & {\color{gelb}$0$} & {\color{gelb}$0.08$} & {\color{gelb}$0.08$}\\ \hline
			\RUw & {\color{gelb}$-0.52$} & {\color{gelb}$0.27$} & {\color{gelb}$-0.44$} & {\color{gelb}$0.54$} & {\color{gelb}$0.02$} & {\color{gelb}$0.09$} & {\color{gelb}$0.31$}\\ \hline
			{\color{bl}\MIik} & {\color{bl}$0.97$} & {\color{bl}$0.99$} & {\color{bl}$1$} & {\color{bl}$1$} & {\color{gelb}$0.64$} & {\color{bl}$0.99$} & {\color{bl}$1$}\\ \hline
		\end{tabular}
		\label{tab:eb_experiment}
	\end{center}
\end{table}

Finding the appropriate behavior of a privacy measure $\PM(x,y)$ is challenging without clear definitions of the private information to be protected. 
We overcome this challenge by specifying requirements for specific scenarios, such as 'Privacy must be minimal for $x=y$.'.
We evaluate privacy measures against these requirements using experiments and tests, grouped by behavior examined: monotonicity, symmetry, and boundary cases. Table~\ref{tab:eb_experiment} presents our findings, which we discuss at the end of this section, with rows aligned with Table~\ref{tab:general-measures-references} and columns explained as we proceed. We will explain the last row, \MIik, in Section~\ref{section:estimator}.

We represent the value of privacy measure $\PM$ as $\prm$. 
The superscript indicates the specific experiment, and the subscript specifies the user load and the grid load that are inputs to the measure. 
For instance, $\prm^\textrm{in}_\textit{ik}$ represents $\PM(x^i,y^i(k))$ in the experiment with user load \textbf{in}terpolation.

\subsection{Monotonicity}
\label{section:monotonicity}
Experiments on monotonicity assume that, as the difference between user load $x$ and grid load $y$ increases, the privacy of a grid load increases as well.
We propose three methods to manipulate the difference between $x$ and $y$: adding noise, interpolating, and compressing $x$.

\subsubsection{Add Noise}
\label{section:AN}
A privacy measure should demonstrate an improvement in privacy as the amplitude $k$ of the noise $u(k)$ added to $x^i$ increases. 
We define $\prm^\textrm{an}_{ik}$ as $\PM(x^i, x^i \oplus u(k))$. 
Then we calculate $\scor^\textrm{an}_{i}$, the Spearman rank correlation between $\prm^\textrm{an}_{ik}$ and $k$ for each user load. 
For two amplitudes ${k_1}$ and ${k_2}$ with ${k_1} > {k_2}$, we expect $\prm^\textrm{an}_{ik_1} \ge \prm^\textrm{an}_{ik_2}$, and therefore $\scor^\textrm{an}_{i}\approx 1$.  
Column AN in Table~\ref{tab:eb_experiment} reports the average correlation $1/N\cdot\sum_{i}\scor^\textrm{an}_i$.

\subsubsection{Interpolation} 
\label{section:IN}
Interpolation involves replacing some user load observations $x^i$ with estimated values to obtain $y^i(k)$
Parameter $k$ controls the granularity of the interpolation, i.e., the number of replaced and interpolated user load values. 
As $k$ increases, a privacy measure should indicate an increase in user privacy.
We define $\prm^\textrm{in}_{ik}=\PM(x^i,y^i(k))$.
For two values ${k_1}$ and ${k_2}$ with $k_1>k_2$, we expect $\prm^\textrm{in}_{ik_1}\ge\prm^\textrm{in}_{ik_2}$.
As before, we calculate the Spearman rank correlation $\scor^\textrm{in}_i$ between $\prm^\textrm{in}_{ik}$ and $k$ for each user load and expect this correlation to be around one. 
Column IN in Table~\ref{tab:eb_experiment} reports the average correlation $1/N\cdot\sum_{i}\scor^\textrm{in}_i$.

\subsubsection{Compression}
\label{section:C}
We use wavelet decomposition to compress $x^i$, where the compression rate depends on the parameter $k$. 
Higher values of $k$ correspond to higher rates.
We calculate the Spearman rank correlation $\scor^\textrm{c}_{i}$ between $\prm^\textrm{c}_{ik}$ and $k$, where $\prm_{ik}^\textrm{c}=\PM(x^i,y^i(k))$.
As the compression rate increases, we expect an increase in privacy and anticipate $\scor^\textrm{c}_{i}$ to be around 1.
Column~C in Table~\ref{tab:eb_experiment} reports $1/N\cdot\sum_{i}\scor^\textrm{c}_i$.

\subsection{Symmetry}
\label{subsection:symmetry}
Symmetry in privacy measures is desirable, as it ensures $\PM(x,y)=\PM(y,x)$. 
Asymmetric measures can produce counterintuitive and unexpected results, e.g., when a measure does not depend on user load or grid load.
While a formula of a privacy measure often indicates whether it is symmetric or not, it does not provide a way to quantify the degree of asymmetry one can expect in real-world scenarios. 
To estimate asymmetry, for each pair of user loads $x^i$ and $x^j$ with $i\ne j$ and a privacy measure $\PM$, we compute $\prm^\textrm{s}_\textit{ij}=\PM(x^i,x^j)$.
Then we compute the ratio $\textit{sym}_k=\sigma_\textit{ik}/\sigma_\textit{kj}$.
Here, $\sigma_\textit{kj}$ is the standard deviation of $\prm^s_\textit{ij}$ when $i=k$, $\sigma_\textit{kj}$ is the standard deviation of $\prm^\textrm{s}_\textit{ij}$ when $j=k$. 
A symmetric privacy measure has $\textit{sym}_k=1$, while $\textit{sym}_k>1$ indicates that changes in $x$ have a greater impact on $\PM(x,y)$ than changes in $y$.
Column SY in Table~\ref{tab:eb_experiment} reports $1/N\cdot\sum_{k=1}^N \textit{sym}_k$.

\subsection{Boundary Cases}
These tests aim at evaluating performance of the privacy measures in extreme cases of maximum or minimum privacy.

\subsubsection{Best Privacy}
\label{section:BP}
The highest level of privacy should be attained by replacing the user load with noise, regardless of the amplitude of the noise.
For each user load $x^i$, we calculate $\prm^\textrm{best}_{ik} = \PM(x^i, \bar{x}^i\oplus u(k))$, 
where $u(k)$ is a noise profile, as defined in Section~\ref{section:notations}, and $\bar{x}^i$ is a constant load profile so that $\bar{x}_t^i=\sum_{l=1}^{T}x^i_l/T$. Parameter $k$ controls the amplitude $k$ of the noise $u(k)$, as in Section~\ref{section:AN}.

To check whether the amplitude of noise has no effect on the privacy estimate,
we compute a p-value for the significance test of the Spearman rank correlation between $\prm^\textrm{best}_\textit{ik}$ and the noise amplitude $k$. 
Define $\pval^\textrm{best}_{i}$ that equals 0 if the p-value is less than 0.1 and 1 otherwise.
If $\prm^\textrm{best}_\textit{ik}$ does not depend on $k$, then $\pval^\textrm{best}_{i}=1$.
Column BP\textsubscript{1} in Table~\ref{tab:eb_experiment} reports $1/N\cdot\sum_{i}\pval^\textrm{best}_i$.

In the second test, we assess whether a privacy measure indicates the highest level of privacy when noise replaces a uses load. %
For each privacy measure and load profile $x^i$, we calculate the proportion of experiments $\textit{sh}^\textrm{best}_i$ described in Section~\ref{section:monotonicity} where the privacy measure indicates lower privacy compared to any of $\prm^\textrm{best}_\textit{ik}$.
If a privacy measure complies with the requirement, we expect $\textit{sh}^\textrm{best}$ to approach~$1$.
Column BP\textsubscript{2} in Table~\ref{tab:eb_experiment} reports $1/N\cdot\sum_{i}\textit{sh}^\textrm{best}_i$.

\subsubsection{Worst Privacy}
\label{section:LP}
If user load remains unchanged, a measure should indicate minimal privacy.
To test this, we use the outcome of all experiments described so far. 
For every load profile $x^i$ and a privacy measure $\PM$, we compute the share of experiments $\textit{sh}^\textrm{worst}_i$ where the privacy measure indicated higher privacy than $\PM(x^i,x^i)$.
If the requirement is met, we expect $\textit{sh}^\textrm{worst}_i$ to be close to one. 
This means that privacy is minimal if one does not modify the user load.
Column LP in Table~\ref{tab:eb_experiment} reports the mean of this share, $1/N\cdot\sum_{i}\textit{sh}^\textrm{worst}_i$.

\subsection{Results}

Table~\ref{tab:eb_experiment} presents the results of these tests, for CER data. 
In each column, values closer to 1 indicate better performance in the corresponding test. 
For better readability, we use black for values within the range $[0.9,1.1]$ and gray for the rest. 
We consider measures within this range to pass the respective test; and the range choice does not impact the ranking of privacy measures much.
Among the 25 examined measures, only five pass at least six out of seven tests.
These measures belong to three groups: coefficient of determination, conditional entropy, and mutual information.

\section{Measure Consistency}
\label{section:methodology_consistency}

\begin{figure}[t]%
	\centering
	\includegraphics[width=\columnwidth]{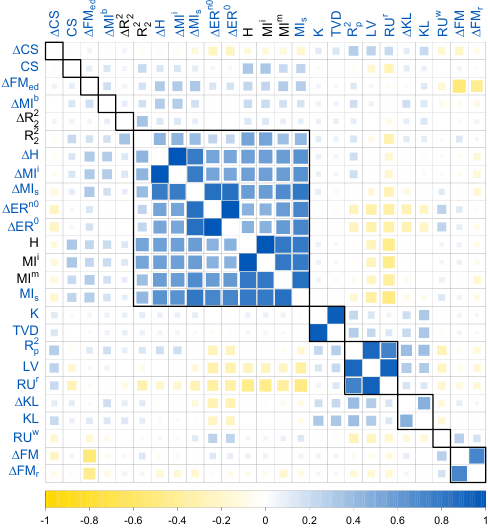}
	\caption{Consistency of privacy measures on CER data}
	\label{fig:corr}
\end{figure}

In the previous section we examined the behavior of individual measures. 
In this section, we investigate the consistency of pairs of measures.
Let $y'$ and $y''$ be two grid loads that correspond to a user load $x$, and $\Delta\PM=PM(x,y')-\PM(x,y'')$.
We say that measures $\PM_1$ and $\PM_2$ are consistent if
$\operatorname{sgn}\Delta\PM_1 = \operatorname{sgn}\Delta\PM_2$ for any choice of $x$, $y'$, and $y''$.
If a pair of measures is consistent, one can use either one (or neither) to compare the performance of load hiding algorithms.

For a pair of privacy measures $\PM_1$ and $\PM_2$, we calculate $\prm'_\textit{ij}=\PM_1(x^i, x^j)$ and $\prm''_\textit{ij}=\PM_2(x^i, x^j)$ for each pair of load profiles $x^i$ and $x^j$.
Then we compute the Spearman rank correlation coefficient $\scor_i$ between $\prm'_\textit{ij}$ and $\prm''_\textit{ij}$ for each $x^i$, $i\in{1,\dots,N}$, and take the average of all $\scor_i$ values to obtain $\scor={1}/{N}\cdot\sum_{i=1}^N \scor_i$.
If the measures are consistent, we expect $\scor$ to equal one. 

Figure~\ref{fig:corr} shows the experiment results, with shaded squares in a cell representing the degree of consistency between measures, and dark blue indicating a high consistency of $p\approx1$. 
We group the measures in each plot based on their degree of consistency. 
We highlight the top five measures from Table~\ref{tab:eb_experiment} in black. 
We observe strong consistency for four pairs of measures: \dERnz and \dERz ($p>0.99$), \MIi and \CE ($p=1$), \dMIi and \dCE ($p=1$), and \K and \TVD ($p>0.95$).
Among the top five measures,  only \MIi and \CE exhibit high consistency.
The low consistency of other top-ranked measures indicates the need for further investigation to select the most appropriate measure, as we discuss in the next section.

\section{Experiments with Synthetic Data}
\label{section:case_study} 

In this section, we use synthetic data from Section~\ref{section:data} to investigate the effectiveness of privacy measures in detecting the presence or absence of secrets in a user load $x$ and grid load $y$. 
The focus is on the top five privacy measures from Table~\ref{tab:eb_experiment}: \Rtwo, \MIi, \MIm, \dRtwo, and \CE. 
To avoid negating the values of \Rtwo, \MIi, \MIm, and \dRtwo, in what follows, we assume that a higher value of $\PM(x,y)$ indicates lower privacy.

\subsection{Relationship of Measures with Secrets}
\label{subsection:RPMS}

\begin{table}[t]%
	\caption{Privacy measures' values for Algorithms~A--D.}
	\begin{center}
	\begin{tabular}{|c|c|c|c|c|c|}
		\hline
		 & \Rtwo & \MIi & \MIm & \dRtwo & \CE \\
		\hline
		A & 0 & 0.04 & 0.04 & 0 & -2.97 \\ \hline
		B & 0.88 & 2 & 0.01 & 0.07 & -1.61 \\ \hline
		C & 0.07 & 2 & 0.01 & 0.01 & -1.61 \\ \hline
		D & 0.51 & 2 & 0.01 & 0 & -1.61 \\
		\hline
	\end{tabular}
\label{tab:mii_rtwo}
\end{center}
\end{table}

Using Algorithms A--D, we generated user and corresponding grid loads of length $T=6400$ with a measurement frequency of $f=200$ readings per day, equivalent to a month of observations. 
Then we applied the top five privacy measures to estimate the privacy achieved. 
Let $\prm^A$, $\prm^B$, $\prm^C$, and $\prm^D$ represent the privacy achieved by applying Algorithm A--D to the user load $x$, respectively. 
Since higher values of $\PM(x,y)$ imply lower privacy, a useful privacy measure according to Section~\ref{section:data} should satisfy: $\prm^A<\prm^B$, and $\prm^A\le\prm^D\le\prm^C\le\prm^B$.

Table~\ref{tab:mii_rtwo} features the results. 
We see that \MIm and \Rtwo fail to meet the inequalities, while \MIi and \CE are suitable when the algorithm is known to an adversary, while
\dRtwo might be preferred otherwise. 
In Section~\ref{section:methodology_consistency}, we found \MIi and \CE to be interchangeable.
Table~\ref{tab:mii_rtwo} suggests that \MIi is more convenient since it approaches 0 for perfect privacy and is independent of $x$. 
So we do not consider \CE any further.

\subsection{Dependence of Measurement Frequency}
\label{subsection:convergence}

\begin{table}[t]%
	\caption{\dRtwo contingent on frequency $f$, $T=6400$.}
	\begin{center}
	\begin{tabular}{|c|c|c|c|c|c|c|c|c|}
		\hline
		$f$ & \textbf{4} & \textbf{8} & \textbf{20} & \textbf{40} & \textbf{100} & \textbf{200} & \textbf{400} & \textbf{800}\\
		\hline
		\dRtwo & 0.9 & 0.81 & 0.61 & 0.41 & 0.18 & 0.07 & 0.02 & 0.01\\ \hline
	\end{tabular}
\label{tab:converge2}
\end{center}
\end{table}

Some measures, e.g., with $\Delta$ notation, make use of the temporal ordering of smart meter measurements.
The measurement frequency may affect these measures, as it changes the expected difference between consecutive measurements. 
We analyzed the effect of measurement frequency on \dRtwo for load profiles $x$ and $y$ corresponding to Algorithm B. 
Table~\ref{tab:converge2} shows that \dRtwo is sensitive to measurement frequency, approaching 0 for high-frequency load profiles. 
This indicates that \dRtwo may not be reliable for detecting whether the 'appliances usage' secret is hidden, as we have expected in Section~\ref{subsection:RPMS}.

\subsection{Sensitivity to Hyperparameters}
\label{subsection:stability}

\begin{table}[t]%
	\caption{\MIm and the shift of ``features'', $f=200$, $T=6400$.}
	\begin{center}
	\begin{tabular}{|c|c|c|c|c|c|c|c|}
		\hline		
		\textbf{shift} & \textbf{0} & \textbf{1} & \textbf{2} & \textbf{4} & \textbf{8} & \textbf{16} & \textbf{25}\\
		\hline
		\MIm & 0.01 & 0.16 & 0.26 & 0.42 & 0.65 & 0.92 & 1.01\\
		\hline
	\end{tabular}
\label{tab:mim}
\end{center}
\end{table}

Hyperparameters may play a crucial role with some privacy measures. 
For example, \Rtwo and \dRtwo require a threshold for the number of timestamps to align user and grid loads~\cite{DBLP:conf/eenergy/ArzamasovSKBN20}. 
However, little research has been done to understand the sensitivity of privacy measures to hyperparameter values.
Our experiments with synthetic data showed stability in \Rtwo and \dRtwo, 
and we omit these results for brevity. %
The ``features'' hyperparameter in \MIm is a vector of the same length as a load profile ($T$), designed to segment a load profile into homogeneous parts~\cite{Chin2018}.
In Section~\ref{subsection:RPMS}, 
we instantiated ``features'' with the ID of the active device,
which required knowing of the ``appliances usage'' secret. %
To investigate the impact of ``features'' on \MIm, we shifted ``features'' by several time stamps and computed \MIm values for different shifts. 
Table~\ref{tab:mim} shows the results for Algorithm~B and indicates that \MIm is very sensitive to the delay, making it challenging to set ``features'' in practice.

\subsection{Summary}
We evaluated the effectiveness of potential privacy measures, \Rtwo, \MIi, \MIm, \dRtwo, and \CE, to detect whether a grid load hides the ``appliances usage''. %
Our results suggest that \MIi and \CE are effective measures for this purpose, and \MIi is more convenient due to its more interpretable values.

\section{Impact of Estimator Choice}
\label{section:estimator}

\begin{table}[t]%
	\caption{Median \MIi depending on hyperparameter value.}
	\begin{center}
	\begin{tabular}{|c|c|c|c|c|c|c|}
		\hline
		& $k=1$ & $k=2$ & $k=4$ & $h=10$ & $h=20$ & $h=40$\\
		\hline
		CER & 0.1 & 0.09 & 0.08 & 0.05 & 0.14 & 0.32\\ \hline
		Smart* & 1.46 & 1.17 & 0.72 & 0.06 & 0.17 & 0.4\\
		\hline
	\end{tabular}
\label{tab:mii_cer_smart}
\end{center}
\end{table}

\begin{table}[t]%
	\caption{Error statistics of \MIi estimators on synthetic data}
	\begin{center}
		\begin{tabular}{|c|c|c|c|c|c|c|}
			\hline
			& $k=1$ & $k=2$ & $k=4$ & $h=10$ & $h=20$ & $h=40$\\
			\hline
			median & 0.02 & 0.04 & 0.06 & 0.56 & 0.28 & 0.03 \\ \hline
			sd & 0.04 & 0.04 & 0.04 & 0.34 & 0.22 & 0.25\\
			\hline
		\end{tabular}
		\label{tab:accuracy_mii2}
	\end{center}
\end{table}

In Section~\ref{subsection:stability}, we noted that many privacy measures have user-specified hyperparameters. 
Other measures rely on hyperparameters through the choice of an estimator. 
For example, \MIi, KL-divergence, entropy ratio, and conditional entropy use the histogram estimator with $h=20$ bins, according to literature on smart meter privacy~\cite{DBLP:conf/eenergy/ArzamasovSKBN20}.
However, the nearest-neighbor (NN) estimator, which requires specifying the number of nearest neighbors $k$ (recommended to be 2--4), may be more data-efficient and less biased~\cite{PhysRevE.69.066138}. 
The suitability of this estimator for smart meter privacy has not yet been explored. 
Our experiments compare these two estimators for \MIi, which has shown promising results so far.

We experiment with real and synthetic load profiles.
For real data, we test the NN estimator with $k=2$ against natural requirements. 
The last row of Table~\ref{tab:eb_experiment} shows that it passes six out of seven tests.
Next, we leverage the design from Section~\ref{section:methodology_consistency} and compute \MIi between pairs of load profiles using the histogram and NN estimators with different values of hyperparameters. 
Table~\ref{tab:mii_cer_smart} shows that the choice of the estimator significantly affects the estimated \MIi value. 
Both estimators have drawbacks: 
The histogram estimator is more sensitive to the choice of $h$ than the NN estimator to the choice of $k$. 
However, the NN estimator has produced negative values in 5--10\% of cases, which are not meaningful.

Knowing the true distribution of synthetic data allows us to calculate the true value of \MIi. 
By manipulating user vs grid load plots, one can control the true \MIi value~\cite{mic}. 
We generate 100 pairs of user and grid loads with varying lengths and \MIi values, and estimate \MIi with histogram and NN estimators. Tables~\ref{tab:accuracy_mii2} shows the error statistics for different values of $k$ and $h$ and indicates that the NN estimator is more accurate and stable regarding its hyperparameter than the histogram estimator.

The NN estimator outperformed the histogram variant with synthetic data, but comparing them on real data did not yield definitive conclusions. %
The choice of estimator greatly impacts measurement results, necessitating further investigation. %

\section{Conclusions}
We studied 25 privacy measures for load hiding algorithms proposed in the literature, to determine the most effective one(s). 
We designed and carried out experiments to assess their compliance with natural requirements, resulting in the elimination of 20 measures deemed ineffective. 
Among the remaining five measures, we found two to be interchangeable. 
We then compiled a list of secrets detectable in smart meter measurements and tested various measures with synthetic data to determine the best one to deal with the ``appliance usage'' secret. 
We find a variant of mutual information (\MIi) to be the most effective measure. 
Interestingly, \MIi performed better in our experiments than other variants of mutual information, such as \MIs and \MIm, which have been proposed in the literature as improvements of \MIi.  
For \MIi, we explored which estimator to use. 
This has received little attention in the literature so far. 
We found drawbacks of the commonly used histogram estimator, and the nearest neighbor estimator may be better. 
All in all, our approach can serve as a blueprint for future studies of other secrets.

\bibliographystyle{IEEEtran}  
\bibliography{Mendeley_Privacy}

\end{document}